\documentclass[fleqn,twoside]{article}
\usepackage[headings]{espcrc2}

\readRCS
$Id: espcrc2.tex,v 1.2 2004/02/24 11:22:11 spepping Exp $
\usepackage{graphicx}

\newcommand{\AmS}{{\protect\the\textfont2
  A\kern-.1667em\lower.5ex\hbox{M}\kern-.125emS}}
\newcommand{\aff}[2]{Dipartimento di Fisica dell'Universit\`a #1 e Sezione INFN, #2, Italy.}
\newcommand{\affd}[1]{Dipartimento di Fisica dell'Universit\`a e Sezione INFN, #1, Italy.}

\title{Search for $\phi\rightarrow K^0\overline{K^0}\gamma$ decay with KLOE}

\author{The KLOE Collaboration: 
F.~Ambrosino\address[Na]{Dipartimento di Scienze Fisiche dell'Universit\`a ``Federico II'' e Sezione INFN, Napoli, Italy},
A.~Antonelli\address[Frascati]{Laboratori Nazionali di Frascati dell'INFN, Frascati, Italy.},
M.~Antonelli\addressmark[Frascati],
F.~Archilli\address[Roma2]{\aff{``Tor Vergata''}{Roma}},
C.~Bacci\address[Roma3]{\aff{``Roma Tre''}{Roma}},
P.~Beltrame\address[Karlsruhe]{Institut f\"ur Experimentelle Kernphysik, Universit\"at Karlsruhe, Germany.},
G.~Bencivenni\addressmark[Frascati],
S.~Bertolucci\addressmark[Frascati],
C.~Bini\address[Roma1]{\aff{``La Sapienza''}{Roma}},
C.~Bloise\addressmark[Frascati],
S.~Bocchetta\addressmark[Roma3],
V.~Bocci\addressmark[Roma1],
F.~Bossi\addressmark[Frascati],
P.~Branchini\addressmark[Roma3],
R.~Caloi\addressmark[Roma1],
P.~Campana\addressmark[Frascati],
G.~Capon\addressmark[Frascati],
T.~Capussela\addressmark[Na],
F.~Ceradini\addressmark[Roma3],
S.~Chi\addressmark[Frascati],
G.~Chiefari\addressmark[Na],
P.~Ciambrone\addressmark[Frascati],
E.~De~Lucia\addressmark[Frascati],
A.~De~Santis\addressmark[Roma1],
P.~De~Simone\addressmark[Frascati],
G.~De~Zorzi\addressmark[Roma1],
A.~Denig\addressmark[Karlsruhe],
A.~Di~Domenico\addressmark[Roma1],
C.~Di~Donato\addressmark[Na],
S.~Di~Falco\address[Pisa]{\affd{Pisa}},
B.~Di~Micco\addressmark[Roma3],
A.~Doria\addressmark[Na],
M.~Dreucci\addressmark[Frascati],
G.~Felici\addressmark[Frascati],
A.~Ferrari\addressmark[Frascati],
M.~L.~Ferrer\addressmark[Frascati],
G.~Finocchiaro\addressmark[Frascati],
S.~Fiore\addressmark[Roma1]
C.~Forti\addressmark[Frascati],
P.~Franzini\addressmark[Roma1],
C.~Gatti\addressmark[Frascati],
P.~Gauzzi\addressmark[Roma1],
S.~Giovannella\addressmark[Frascati],
E.~Gorini\address[Lecce]{\affd{Lecce}},
E.~Graziani\addressmark[Roma3],
M.~Incagli\addressmark[Pisa],
W.~Kluge\addressmark[Karlsruhe],
V.~Kulikov\address[Moscow]{Permanent address: Institute for Theoretical and Experimental Physics, Moscow, Russia.},
F.~Lacava\addressmark[Roma1],
G.~Lanfranchi\addressmark[Frascati],
J.~Lee-Franzini\addressmark[Frascati]\address[StonyBrook]{Physics Department, State University of New 
York at Stony Brook, USA.},
D.~Leone\addressmark[Karlsruhe],
M.~Martini\addressmark[Frascati],
P.~Massarotti\addressmark[Na],
W.~Mei\addressmark[Frascati],
S.~Meola\addressmark[Na],
S.~Miscetti\addressmark[Frascati],
M.~Moulson\addressmark[Frascati],
S.~M\"uller\addressmark[Frascati],
F.~Murtas\addressmark[Frascati],
M.~Napolitano\addressmark[Na],
F.~Nguyen\addressmark[Roma3],
M.~Palutan\addressmark[Frascati],
E.~Pasqualucci\addressmark[Roma1],
A.~Passeri\addressmark[Roma3],
V.~Patera\addressmark[Frascati]\address[Energ]{Dipartimento di Energetica dell'Universit\`a ``La Sapienza'', Roma, Italy.},
F.~Perfetto\addressmark[Na],
M.~Primavera\addressmark[Lecce],
P.~Santangelo\addressmark[Frascati],
G.~Saracino\addressmark[Na],
B.~Sciascia\addressmark[Frascati],
A.~Sciubba\addressmark[Frascati]\addressmark[Energ],
F.~Scuri\addressmark[Pisa],
I.~Sfiligoi\addressmark[Frascati],
T.~Spadaro\addressmark[Frascati],
M.~Testa\addressmark[Roma1],
L.~Tortora\addressmark[Roma3],
P.~Valente\addressmark[Roma1],
B.~Valeriani\addressmark[Karlsruhe],
G.~Venanzoni\addressmark[Frascati],
R.Versaci\addressmark[Frascati],
G.~Xu\addressmark[Frascati]\address[Beijing]{Permanent address: Institute of High Energy Physics of Academica Sinica,  Beijing, China.}}

\runtitle{Search for $\phi\rightarrow K^0\overline{K^0}\gamma$ decay with KLOE}
\runauthor{The KLOE Collaboration}

\begin{document}

\begin{abstract}
The KLOE collaboration has searched for the $\phi\rightarrow K^0\overline{K^0}\gamma$ decay using a sample of 1.4 fb$^{-1}$ of  $e^+$ $e^-$ collisions at $\sqrt{s}\sim M(\phi)$ collected with the KLOE experiment at the Frascati $e^+$ $e^-$ collider DA$\Phi$NE. No previous search exists for this decay, while many theory models predict a BR of $\approx10^{-8}$ for this channel. We set a preliminary value of the U.L. on this BR to $1.8\cdot10^{-8}$ at 90\% C.L.. This limit rules out most of the existing theory predictions.
\vspace{1pc}
\end{abstract}

\maketitle

\section{Introduction}
We present the results of a search for the decay 
$\phi\rightarrow K^0\overline{K^0}\gamma$ using 1.4 fb$^{-1}$ of the KLOE data sample. This
decay has never been searched before. The $\phi$ resonance is produced through e$^+$e$^-$ collisions at center of mass energy $\sqrt{s}\sim$1020 MeV. In this decay the $K^0\overline{K^0}$
pair is produced with positive charge conjugation, so that the state of the two kaons can be described as
\begin{equation}
|K^0\overline{K^0}>={{|K_SK_S>+|K_LK_L>}\over{\sqrt{2}}} .
\end{equation}
The signature of this decay is provided by the presence of either 2
$K_S$ or 2 $K_L$ and a low energy photon. This process has a limited phase space due to the small difference between the $\phi$ mass (1019.5 MeV) and the production threshold of two neutral kaons (995 MeV). This results in a very narrow photon energy spectrum, ranging 
from 0 up to 
a maximum energy
obtained when the two kaons are collinear and the KK 
invariant mass is equal to twice the kaon mass, that is (neglecting the small $\phi$ momentum due to the e$^+$e$^-$ crossing angle): 
\begin{equation}
E_{\gamma,max}={{M^2_{\phi}-(2M_K)^2}\over{2M_{\phi}}}=23.8~{\rm MeV}
\end{equation}
Among the possible final states, we searched for that one where a $K_SK_S$ pair has both $K_S$ decaying to $\pi^+\pi^-$. This corresponds to reduce the 
rate of the searched events of a fraction
\begin{equation}
{{1}\over{2}}\times B.R.(K_S\rightarrow \pi^+\pi^-)^2=23.9\%
\end{equation}
This decay chain is characterized by a clean signature: 2 vertices close to the
interaction region, with both vertices having two tracks with opposite sign, an invariant mass equal to the kaon mass, and an invariant mass of the 2 kaons significantly lower
than the $\phi$ mass. Moreover, a low energy photon should be present in the
event.  

\begin{figure}[htbp]
  \centerline{
    \resizebox{1.0\columnwidth}{1.0\columnwidth}{
      \includegraphics{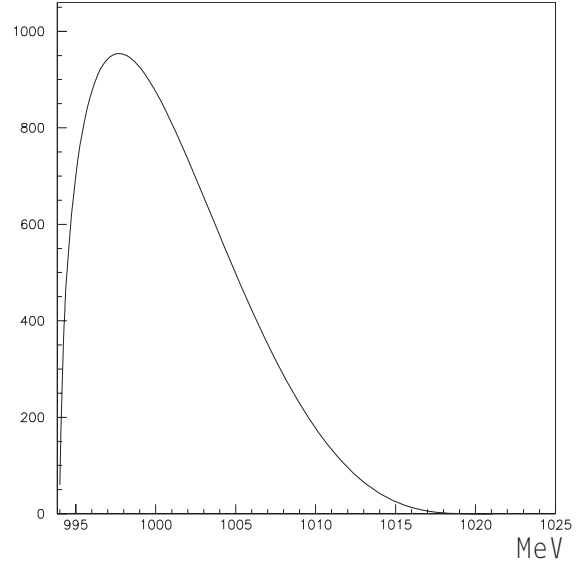}
    }
  }
\caption{Expected scalar meson's invariant mass spectrum according to phase-space and radiative decay dinamics.}
\label{fig:1}
\end{figure}


\par In the paper, we first discuss the
motivations of this search, we then describe the analysis method including the
Montecarlo study, and we conclude by extracting the upper limit on the branching ratio and compare it with the theory expectations.

\section{Motivations}
The $\phi\rightarrow K^0\overline{K^0}\gamma$ process was considered in the
KLOE proposal as a possible background source for the CP violation
measurement. The conclusion was that only 
for branching ratios in excess of $\sim
10^{-6}$ such a background could be critical for the measurement if no
selection on the photon and on the kinematics was applied. 
\par On the other
hand the value of the branching ratio gives relevant information on the scalar
mesons structure. The $K^0\overline{K^0}$ state can have scalar quantum
numbers in both triplet and singlet isospin state, so that 
the reaction is expected to proceed mainly through
the chain $\phi\rightarrow ({\rm f_0(980)+a_0(980)})\gamma\rightarrow
K^0\overline{K^0}\gamma$. The prediction on the branching ratio depends on the
way the scalar dynamics is introduced and on the size of the couplings of the
scalars to the kaons. Interference effects between f$_0$ and a$_0$ amplitudes
can also be present.
\par
Theory predictions on the BR($\phi\rightarrow K^0\overline{K^0}\gamma$)  found in literature spread over several orders of magnitude. The latest
evaluations 
essentially concentrate in the region of $10^{-8}$. All of them are
well below the critical limit of $10^{-6}$ so that no significant effect is
expected for the CP violation studies at a $\phi$-factory. 
\par 
Some of the reported predictions do not include explicitely the
scalar mesons, but consider them as dynamically
generated in the theory \cite{Bramon2,Oller,Escribano}; most of the theory instead includes explicitely the scalars mesons\cite{Achasov2,Gokalp,Achasov1,Nussinov,Lucio,Close,Pacetti} in the calculation of the BR, in such a way that the predicted value depends on this modeling. For instance, the two predictions of ref.6 are evaluated assuming a 2-quark or a 4-quark struture for the scalar mesons. 
in such a way that
the predicted value depends on the way they are treated. The two predictions
differ by one order of magnitude.

In the other cases the width of the allowed
band is due to the uncertainty on the 
coupling constants used in the parametrization of the amplitude which is extracted by experimental analysis of $\phi\rightarrow \pi\pi\gamma$
and $\phi\rightarrow \eta\pi\gamma$. 
The latter approach is particularly interesting, since it allows to make a
global analysis of KLOE data including $\pi\pi\gamma$ and $\eta\pi\gamma$\cite{pl1,pl2,pl3,epjc1} to test consistency of the overall picture.

Other predictions\cite{Fajfer,Bramon1} do not include the
scalars so that have to be considered as ``backgrounds'' in the search for
effects due to scalars; 

\section{Experimental setup}
The KLOE experiment is performed at the Frascati $\phi$ factory DA$\Phi$NE, an $e^+$ $e^-$ collider running at $\sqrt{s} \sim 1020$ MeV ($\phi$ mass).
Beams collide with a crossing angle of
($\pi-0.025$) rad. From 2001 to 2005, the KLOE experiment has collected an integrated luminosity of ~2.5 fb$^{-1}$

The KLOE detector consists of
a large-volume cylindrical drift chamber~\cite{dch} (3.3~m length
and 4m diameter),
surrounded by a sampling calorimeter~\cite{calo} made of lead and
scintillating fibres. 
The detector is inserted in a superconducting coil 
producing a solenoidal field $B$=0.52~T.
Large-angle tracks from the origin ($\theta>45^\circ$) are
reconstructed with momentum resolution $\sigma_p/p=0.4\%$.
Photon energies and times are measured by the calorimeter with
resolutions $\sigma_E/E=5.7\%/\sqrt{E({\rm GeV})}$ and 
$\sigma_t=54\mbox{ ps}/\sqrt{E({\rm GeV})}\oplus 50$ ps.

\section{Analysis strategy}

The event selection performed by this analysis is based on kinematic cuts on the charged pion tracks detected by the drift chamber, and on the photon cluster identification in the calorimeter.

We analysed 1.4 fb$^{-1}$ of data collected at the $\phi$ peak; we also used our Monte Carlo (MC) to generate an equivalent statistics of background, which is mainly due to $\phi \rightarrow K_SK_L \rightarrow \pi^+\pi^-\pi^+\pi^-$ with the $K_L$ decaying close to the IP and an additional ISR, FSR photon. Our simulation is generated on a run-by-run basis, using as input the real data taking conditions for both detector and collider.

We have also generated a MC signal sample of 10k events, to study the analysis selection efficiency. The main ingredient of this simulation is the scalar meson's invariant mass shape, which slightly depends on the scalar meson structure. We did not use any of the models quoted above, but instead relied on general assumptions of BR dependence on the radiated photon's energy and on phase space. In Fig.\ref{fig:1} the generated invariant mass spectrum is shown.

We look for two $K_S$ decaying into charged pions, by requiring the presence of two vertices close to the interaction point, inside a fiducial volume defined as a cylinder of 3 cm radius in the transverse plane, and $\pm$8 cm along the beam line. Each vertex should have two charged tracks attached to.

\begin{figure}[htbp]
  \centerline{
    \resizebox{1.0\columnwidth}{0.9\columnwidth}{%
      \includegraphics{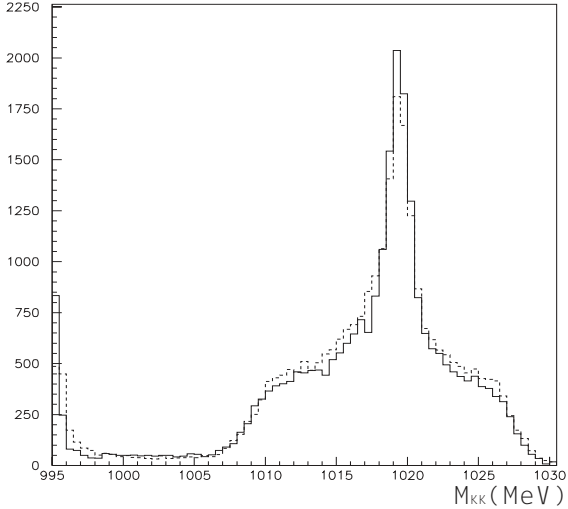}
    }
  }
\caption{Invariant mass of the kaon pair $M_{KK}$ for Monte Carlo (solid) and Data (dashed).}
\label{invmasskk}
\end{figure}

For each vertex, the two track reconstructed mass, $M_{2\pi}$, is built in the pion hypothesys. For the signal, the event density in the $M_{2\pi}(1)$, $M_{2\pi}(2)$ plane is well contained inside a circle of few MeV radius around the $K_S$ mass. We require the events to satisfy a 4 MeV cut on this radius.
With the reconstructed masses and momenta of the two $K_S$ candidates we calculate the invariant mass of the kaon pair.
As expected by the signal simulated mass spectra shown in Fig.\ref{fig:1}, the $M_{KK}$ distribution peaks at 1000 MeV, while the background is peaked at $M_\phi$ as shown in Fig.\ref{invmasskk}. A large background reduction is obtained by rejecting events with $M_{KK} >$ 1010 MeV. Moreover, the 4-momentum conservation in the $\phi \rightarrow K_sK_s\gamma$ decay allows to build a missing 4-momentum $\widetilde{P_\gamma}=\widetilde{P_\phi}-\widetilde{P_{K1}}-\widetilde{P_{K2}}$ based only on track reconstructed variables. A selecion variable $M^2_{\gamma}=E^2_{miss}-P^2_{miss}$ is built, which is expected to be $\approx$ 0 for the signal. We retain events with $|{M^2_{\gamma}}|\leq 300MeV^2$.

Events that survive all these cuts are searched for the presence of one photon matching missing momentum.
We require the presence of one cluster in the calorimeter not associated with any charged track. Cluster's timing must be compatible with a photon coming from the interaction point, and cluster's position and energy must agree within resolution with the missing momentum. 

\section{Results}
{}
All the above mentioned cuts have been decided upon an U.L. maximization based on MC samples. We estimate an efficiency of 20.6\% for the signal, while we find no event surviving in the background MC. When looking at DATA we observe one event. DATA-MC comparison is still under way, so we do not use background evaluation for this preliminary result. However we can set an upper limit based on Poisson statistics without background subtraction at 90\% C.L. to 3.9 events.

\begin{figure}[htbp]
  \centerline{
    \resizebox{1.0\columnwidth}{1.0\columnwidth}{%
      \includegraphics{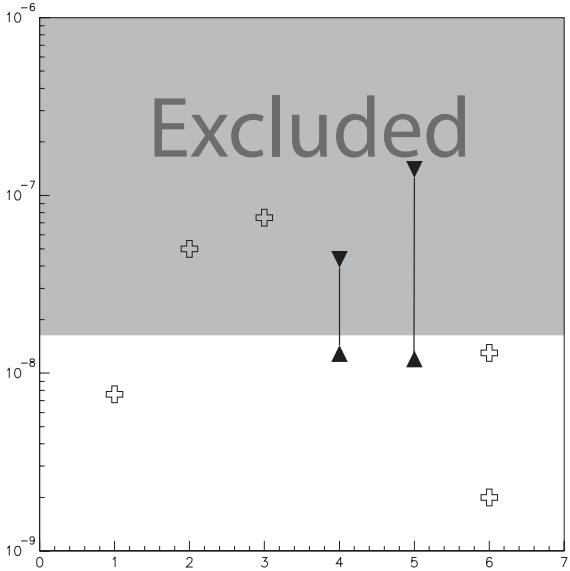}
    }
  }
\caption{Comparison between theoretical predictions and our measurement. In abscissa is directly reported the reference number (according to the reference list). For ref.6 two predictions are reported: the upper one is for 4-quark hypothesys, the lower one for 2-quark hypothesys. For refs.4,5 the prediction is represented as a band . }
\label{predictions}
\end{figure}

We evaluate our B.R. upper limit in the following way:
$BR(\phi \rightarrow K_0\overline{K_0}\gamma) = $
\begin{equation}
\frac{U.L.(\mu_{sig})}{\int{\mathcal{L}dt}\cdot\sigma(e^+e^-\rightarrow \phi)\cdot\frac{1}{2}\cdot(BR(K_S \rightarrow \pi^+\pi^-))^2 \cdot\epsilon}
\end{equation}
in which $\mathcal{L}dt$ is 1.4 fb$^{-1}$, the factor 1/2 accounts for the fact that we are selecting only the K$_S$K$_S$ combination for $K_0\overline{K_0}$ , $\epsilon$ is our signal efficiency
The following limit on the branching ratio is obtained: 
\begin{equation}
BR(\phi \rightarrow K_0\overline{K_0}\gamma) < 1.8\cdot10^{-8}
\end{equation}
at 90\%C.L. In Fig.\ref{predictions} our limit is compared to theoretical predictions. Most of them are excluded by our result.

\end{document}